\documentclass[aps,pre,twocolumn,superscriptaddress,longbibliography,
               amsmath,amssymb,nofootinbib]{revtex4-2}

\usepackage{graphicx}
\usepackage{bm}
\usepackage{xcolor}
\usepackage{url}

\begin{document}

\title{Charge discreteness and the energy efficiency of information erasure\\
in dynamic random-access memory cells}

\author{Takase Shimizu}
\email{takase.shimizu@ntt.com}
\affiliation{Basic Research Laboratories, NTT Inc., Kanagawa, Japan}
\author{Kouki Yamamoto}
\affiliation{Basic Research Laboratories, NTT Inc., Kanagawa, Japan}
\author{Kensaku Chida}
\affiliation{Basic Research Laboratories, NTT Inc., Kanagawa, Japan}
\author{Gento Yamahata}
\affiliation{Basic Research Laboratories, NTT Inc., Kanagawa, Japan}
\author{Katsuhiko Nishiguchi}
\affiliation{Institute for Multidisciplinary Sciences, Yokohama National University, Kanagawa, Japan}

\begin{abstract}
A dynamic random-access memory (DRAM) cell stores information as an integer number of electrons on a capacitor, and whether this discreteness is thermodynamically relevant depends on the competition between the charging energy and thermal fluctuations.
This competition is quantified by the ratio $\kappa$ of the single-electron charging energy to the thermal energy, and here we investigate how $\kappa$ affects the energy efficiency of information erasure in a DRAM cell.
Using a stochastic-thermodynamic model of a DRAM cell, we show that the nonquasistatic heat released during the discharge step is suppressed as $\kappa$ increases, whereas the quasistatic heat of the charge step approaches the Landauer cost.
As a result, the energy efficiency increases monotonically with $\kappa$ and approaches the Landauer limit where the effect of charge discreteness is maximal and the cell is effectively reduced to two charge states.
The parameter $\kappa$ thus connects two thermodynamic regimes: a multilevel single-well memory, whose nonequilibrium initial state prevents quasistatic erasure, and an effective two-level memory that can attain the Landauer limit.
These results identify $\kappa$ as the parameter that controls the fundamental efficiency ceiling of transistor--capacitor memory circuits.
  
\end{abstract}

\maketitle

\textit{Introduction}---The thermodynamics of information establishes fundamental bounds on the energetic cost of computation~\cite{Landauer1961,Bennett1982,Parrondo2015,Wolpert2019}. Its cornerstone, the Landauer principle, states that erasing one bit of information requires heat dissipation no smaller than the corresponding reduction of the memory entropy. This bound has been tested in a variety of small physical systems, including colloidal particles, feedback traps, nanomagnets, and trapped ions~\cite{Berut2012,Jun2014,Hong2016,Yan2018}. Recent theoretical and experimental studies have further clarified finite-time corrections and protocol-dependent refinements of the bound~\cite{Proesmans2020,Lee2022,VanVu2023,Zhen2021,Scandi2022,Oikawa2025}. A central open question, however, is how such thermodynamic bounds are modified by the structural constraints of concrete electronic devices.

\begin{figure}[tb]
\includegraphics[width=\columnwidth]{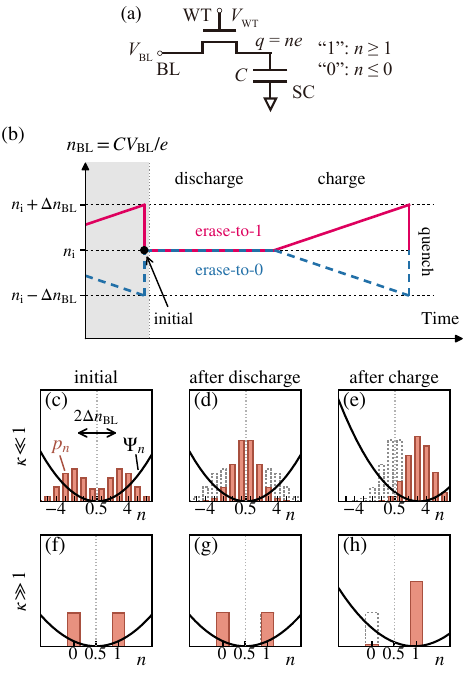}
\caption{\label{fig:scheme}%
Erasure of a DRAM cell and the effect of charge discreteness.
(a)~Equivalent circuit: a storage capacitor (SC, capacitance $C$) connected to a bitline (BL, $V_{\mathrm{BL}}$) through a wordline transistor (WT, $V_{\mathrm{WT}}$); the logical state is determined by the excess number of electrons $n$.
(b)~Bitline-voltage sequences of the erase-to-1 (red) and erase-to-0 (blue dashed) operations: after discharge at $n_{\mathrm{i}}=0.5$, $n_{\mathrm{BL}}$ is swept quasistatically to $n_{\mathrm{i}}+\Delta n_{\mathrm{BL}}$ or $n_{\mathrm{i}}-\Delta n_{\mathrm{BL}}$, respectively, and is then quenched back to $n_{\mathrm{i}}$. The shaded region shows the end of the preceding cycle, whose quench prepares the initial state (marker) from which the discharge step begins.
(c)--(h)~Distribution $p_n$ (bars) and state function $\Psi_n$ (parabola) during an erase-to-1 operation applied to the initial state, for $\kappa=E_{\mathrm{c}}/k_{\mathrm{B}}T\ll 1$ [(c)--(e)] and $\kappa\gg 1$ [(f)--(h)]; dotted outlines show the distribution before each step, and vertical dotted lines mark $n=n_{\mathrm{i}}=0.5$.
For $\kappa\ll 1$ the discharge step relaxes the broad bimodal state through many intermediate charge states, whereas for $\kappa\gg 1$ only $n=0,1$ are populated, so the initial and discharged states coincide and the discharge relaxation is absent.}
\end{figure}

Dynamic random-access memory (DRAM) is a minimal and technologically important example of a transistor--capacitor information-processing circuit. A DRAM cell stores one bit as an integer number of electrons on a storage capacitor [Fig.~\ref{fig:scheme}(a)]. Unlike idealized two-level memories or double-well systems, the state function of a DRAM cell is a single-well parabola in the stored charge, so the logical mixture appearing during memory operation is generally not an equilibrium state of the cell. This nonequilibrium initial state can prevent quasistatic erasure and impose a device-specific efficiency ceiling, as demonstrated experimentally~\cite{Shimizu2026}.

While that experiment demonstrated the existence of such a ceiling for a single device, here we address a more general question: how the discreteness of charge affects the energy efficiency of information erasure---the logical entropy reduction divided by the dissipated heat measured in units of the thermal energy. Although the charge is always discrete, its discreteness is thermodynamically relevant only when the charging energy exceeds the thermal energy. The relevant dimensionless parameter is therefore
\begin{equation}
\kappa = \frac{E_{\mathrm{c}}}{k_{\mathrm{B}}T},
\label{eq:kappa}
\end{equation}
where $E_{\mathrm{c}}=e^{2}/2C$ is the charging energy of a storage capacitor with capacitance $C$, $e$ is the electron charge, $k_{\mathrm{B}}$ is the Boltzmann constant, and $T$ is the temperature. When $\kappa\ll 1$, many integer charge states are thermally populated, and the stored charge can be treated approximately as a continuous variable. In the opposite limit $\kappa\gg 1$, the equilibrium distribution is concentrated on only a few charge states, and the effect of charge discreteness is maximal. Although this crossover is a basic consequence of charge quantization, its impact on the energy efficiency of a DRAM cell has not been clarified.

In this work, we compute the energy efficiency of a DRAM cell over the full range of $\kappa$ using a stochastic-thermodynamic model of the charge dynamics. The erasure protocol consists of a nonquasistatic discharge step followed by a quasistatic charge step, and we evaluate how the heat released in each step changes with $\kappa$. We find that the two steps respond oppositely to $\kappa$: the heat released during discharge is suppressed as $\kappa$ increases, whereas the heat released during the quasistatic charge step approaches the Landauer cost. Consequently, the energy efficiency increases monotonically with $\kappa$ and approaches the Landauer limit where the effect of charge discreteness is maximal and the cell is effectively reduced to two charge states. Thus $\kappa$ determines the energy efficiency of a DRAM cell, bridging two limits: a memory whose charge behaves as a continuous variable, structurally barred from quasistatic erasure, and an effective two-level memory attaining the Landauer limit.

\textit{Model}---We adopt the equivalent-circuit description of a DRAM cell used in Refs.~\cite{Shimizu2026} [Fig.~\ref{fig:scheme}(a)].
The cell stores the charge $q=ne$ ($n$ integer) on the storage capacitor (SC); the logical states are assigned as ``0'' for $n\le 0$ and ``1'' for $n\ge 1$.
The bitline (BL) voltage $V_{\mathrm{BL}}$ sets the target charge of the capacitor and drives the charge step, while the wordline transistor (WT) gate voltage $V_{\mathrm{WT}}$ controls the electron hopping rate between the bitline and the capacitor; throughout this work $V_{\mathrm{WT}}$ is held fixed.
Neglecting the junction capacitance of the WT, the state function of the cell is~\cite{Shimizu2026}
\begin{equation}
\Psi_n(n_{\mathrm{BL}})
 = E_{\mathrm{c}}\left(n-n_{\mathrm{BL}}\right)^{2},
\label{eq:Psi}
\end{equation}
where $n_{\mathrm{BL}}=eV_{\mathrm{BL}}/(2E_{\mathrm{c}})$ is the normalized bitline voltage; an $n$-independent term $-E_{\mathrm{c}}n_{\mathrm{BL}}^{2}$ has been dropped in this study, as it does not contribute to the heat during electron hopping.
Let $p_n$ be the probability that the cell holds $n$ excess electrons. Thermal fluctuations make $n$ vary, and at equilibrium for a given $n_{\mathrm{BL}}$ it follows the discrete Gaussian~\cite{Freitas2021,Nishiguchi2014}
\begin{equation}
p_n^{\mathrm{eq}}(n_{\mathrm{BL}})
 = \frac{\exp\!\left[-\kappa\,(n-n_{\mathrm{BL}})^{2}\right]}
        {\sum_{n'}\exp\!\left[-\kappa\,(n'-n_{\mathrm{BL}})^{2}\right]},
\label{eq:peq}
\end{equation}
of standard deviation $\sigma = 1/\sqrt{2\kappa}$ in the continuum limit.
The parameter $\kappa$ is the central control parameter of this work.

The erasure protocol follows Ref.~\cite{Shimizu2026}, to which we refer for details; the operations are sketched in Fig.~\ref{fig:scheme}(b).
An erase-to-1 operation consists of three steps: (i) a \textit{discharge} step, in which $V_{\mathrm{BL}}$ is held at the voltage $n_{\mathrm{BL}}=n_{\mathrm{i}}=0.5$, the midpoint between the two logical states, until the cell equilibrates; (ii) a \textit{charge} step, in which $n_{\mathrm{BL}}$ is swept quasistatically to $n_{\mathrm{i}}+\Delta n_{\mathrm{BL}}$ ($\Delta n_{\mathrm{BL}}>0$); and (iii) a quench back to $n_{\mathrm{i}}$, which changes neither the charge distribution nor the heat released during the erasure itself~\cite{Shimizu2026}.
Erase-to-0 is defined analogously, except that the charge step ends at $n_{\mathrm{i}}-\Delta n_{\mathrm{BL}}$.
The outcomes of these two operations are therefore $p_n^{\mathrm{eq}}(n_{\mathrm{i}}+\Delta n_{\mathrm{BL}})$ and $p_n^{\mathrm{eq}}(n_{\mathrm{i}}-\Delta n_{\mathrm{BL}})$, respectively.

We now specify the probability distribution reached at each stage of an erase-to-1 operation. The initial state, where the two logical states are equally populated, is prepared as the average of the outcomes of erase-to-0 and erase-to-1 [shaded region in Fig.~\ref{fig:scheme}(b)],
\begin{equation}
p_n^{\mathrm{initial}}
 = \tfrac{1}{2}\!\left[
   p_n^{\mathrm{eq}}\!\left(n_{\mathrm{i}}-\Delta n_{\mathrm{BL}}\right)
 + p_n^{\mathrm{eq}}\!\left(n_{\mathrm{i}}+\Delta n_{\mathrm{BL}}\right)\right],
\label{eq:pinit}
\end{equation}
a bimodal distribution centered at $n = n_{\mathrm{i}} \pm \Delta n_{\mathrm{BL}}$ [Figs.~\ref{fig:scheme}(c) and \ref{fig:scheme}(f)]. Its discharge step relaxes the initial state to the equilibrium distribution at $n_{\mathrm{i}}$,
\begin{equation}
p_n^{\mathrm{discharged}} = p_n^{\mathrm{eq}}(n_{\mathrm{i}})
\label{eq:pdis}
\end{equation}
[Figs.~\ref{fig:scheme}(d) and \ref{fig:scheme}(g)]. Its charge step then prepares the equilibrium distribution at $n_{\mathrm{i}}+\Delta n_{\mathrm{BL}}$,
\begin{equation}
p_n^{\mathrm{final}} = p_n^{\mathrm{eq}}(n_{\mathrm{i}}+\Delta n_{\mathrm{BL}})
\label{eq:pfin}
\end{equation}
[Figs.~\ref{fig:scheme}(e) and \ref{fig:scheme}(h)]. The subsequent quench leaves this distribution unchanged, so $p_n^{\mathrm{final}}$ is also the state at the end of the erasure. The erasure error probability is the residual weight of the ``0'' state in the final distribution, $\varepsilon = \sum_{n\le 0} p_n^{\mathrm{final}}$.

The total heat dissipation is $Q = Q_{\mathrm{discharge}} + Q_{\mathrm{charge}}$, with the two contributions~\cite{Shimizu2026}
\begin{align}
Q_{\mathrm{discharge}}
 &= \sum_n \left[p_n^{\mathrm{initial}} - p_n^{\mathrm{discharged}}\right]
    \Psi_n(n_{\mathrm{i}}),
\label{eq:Qd}\\
Q_{\mathrm{charge}}
 &= k_{\mathrm{B}}T \left[\,s\!\left(p^{\mathrm{discharged}}\right)
                        - s\!\left(p^{\mathrm{final}}\right)\right],
\label{eq:Qc}
\end{align}
where $s(p) = -\sum_n p_n \ln p_n$ is the internal (Shannon) entropy with respect to $n$.
Since the discharge step proceeds at fixed $n_{\mathrm{BL}}=n_{\mathrm{i}}$, the state function $\Psi_n$ does not change and no work is done on the cell, so Eq.~(\ref{eq:Qd}) is the resulting drop in $\langle\Psi\rangle_p=\sum_n p_n\Psi_n(n_{\mathrm{i}})$; it is the nonquasistatic contribution. The charge step is quasistatic, so the cell stays in equilibrium and the dissipated heat equals the decrease of the internal entropy $s(p)$, giving Eq.~(\ref{eq:Qc}).
The logical Shannon entropy of the initial state is exactly $\ln 2$ by the symmetry of Eq.~(\ref{eq:pinit}), and that of the final state is $H(\varepsilon) = -\varepsilon\ln\varepsilon - (1-\varepsilon)\ln(1-\varepsilon)$, so the logical entropy change of the memory is $\Delta S = H(\varepsilon) - \ln 2$.
The energy efficiency is
\begin{equation}
\eta
 := \frac{-\Delta S}{Q/k_{\mathrm{B}}T}
 = \frac{\ln 2 - H(\varepsilon)}
        {(Q_{\mathrm{discharge}} + Q_{\mathrm{charge}})/k_{\mathrm{B}}T},
\label{eq:eta}
\end{equation}
with $\eta = 1$ at the Landauer limit.

\begin{figure}[tb]
\includegraphics[width=\columnwidth]{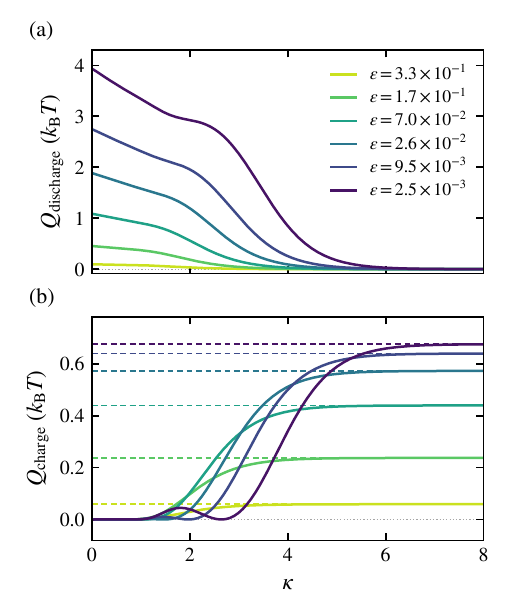}
\caption{\label{fig:heat}%
Calculated heat dissipation in the two steps of the erasure protocol versus $\kappa$, for various values of the erasure error probability $\varepsilon$.
(a)~Discharge step, Eq.~(\ref{eq:Qd}).
$Q_{\mathrm{discharge}}$ decreases monotonically and vanishes for $\kappa \gg 1$.
(b)~Charge step, Eq.~(\ref{eq:Qc}).
$Q_{\mathrm{charge}}$ increases with $\kappa$ and saturates at the Landauer cost $k_{\mathrm{B}}T[\ln 2 - H(\varepsilon)]$, indicated by the dashed horizontal line of the corresponding color for each $\varepsilon$; horizontal dotted lines indicate $Q=0$.}
\end{figure}

\textit{Heat of the discharge and charge steps}---We first examine the heat released in the two steps of the erasure protocol as $\kappa$ is varied. Figure~\ref{fig:heat} shows the results computed from Eqs.~(\ref{eq:Qd}) and (\ref{eq:Qc}) at fixed erasure error probabilities. $Q_{\rm discharge}$ decreases with increasing $\kappa$ and becomes strongly suppressed for $\kappa\gg1$ [Fig.~\ref{fig:heat}(a)]. In contrast, $Q_{\rm charge}$ increases overall and saturates at a finite value for large $\kappa$ [Fig.~\ref{fig:heat}(b)].

The decrease of $Q_{\rm discharge}$ originates from the reduced mismatch between the initial state and the equilibrium state reached after discharge. During the discharge step, $n_{\rm BL}$ is fixed at $n_{\mathrm{i}}$, so heat is released only if the initial state relaxes toward $p^{\rm eq}_n(n_{\mathrm{i}})$. When the charge distribution is broad over many charge states, the two distributions forming the initial state differ substantially from the equilibrium distribution at $n_{\mathrm{i}}$ [Figs.~\ref{fig:scheme}(c) and \ref{fig:scheme}(d)], and the relaxation releases a finite amount of heat. As $\kappa$ increases, the width of each distribution shrinks and, at fixed error probability, the required displacement $\Delta n_{\rm BL}$ also becomes smaller (see Appendix~\ref{app:pinit}). For large $\kappa$, where only $n=0$ and $n=1$ are populated, the initial state nearly coincides with $p^{\rm discharged}$ [Figs.~\ref{fig:scheme}(f) and \ref{fig:scheme}(g)]. The discharge relaxation is then strongly suppressed, so $Q_{\rm discharge}\to0$.

The increase of $Q_{\rm charge}$ reflects the growing internal-entropy difference between the discharged and final distributions in Eq.~(\ref{eq:Qc}). For small $\kappa$, both distributions are broad enough to be well approximated by continuous Gaussians and are nearly translations of each other, so their internal entropies are almost the same. The quasistatic charge step therefore releases little heat. As $\kappa$ increases, the distributions become narrower, and the shift of the final distribution produces a larger internal-entropy reduction. For $\kappa\gg1$, only $n=0$ and $n=1$ are populated, so each logical state maps onto a single charge state; the internal entropy $s(p)$ then equals the logical entropy for every distribution. In particular $s(p^{\rm discharged})\to\ln2$ and $s(p^{\rm final})\to H(\varepsilon)$, so the internal-entropy reduction $s(p^{\rm discharged})-s(p^{\rm final})$ coincides with the logical entropy change $-\Delta S$, and $Q_{\rm charge}$ approaches the Landauer cost $k_{\rm B}T[\ln2-H(\varepsilon)]$.

\begin{figure}[tb]
\includegraphics[width=\columnwidth]{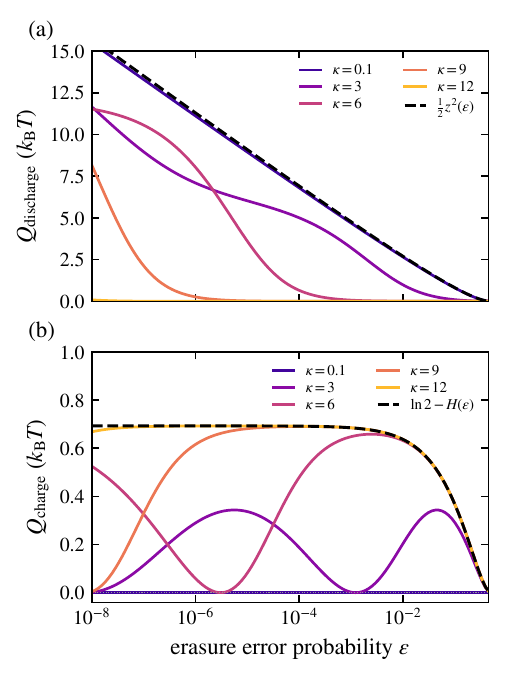}
\caption{\label{fig:heat-eps}%
Heat dissipation versus erasure error probability $\varepsilon$, from the exact evaluation of Eqs.~(\ref{eq:Qd}) and (\ref{eq:Qc}).
(a)~$Q_{\mathrm{discharge}}$ for several values of $\kappa$; the dashed curve is the continuum-limit expression $\tfrac{1}{2}k_{\mathrm{B}}T z^{2}(\varepsilon)$ [Eq.~(\ref{eq:Q0})].
(b)~$Q_{\mathrm{charge}}$ for several values of $\kappa$; the dashed curve is the Landauer cost $k_{\mathrm{B}}T[\ln 2 - H(\varepsilon)]$, approached for $\kappa\gg1$ [Eq.~(\ref{eq:Qc-limit})].}
\end{figure}

Figure~\ref{fig:heat-eps} shows the heat as a function of $\varepsilon$ for various values of $\kappa$. In Fig.~\ref{fig:heat-eps}(a), $Q_{\rm discharge}$ increases as $\varepsilon$ decreases, because a smaller $\varepsilon$ requires a larger $\Delta n_{\rm BL}$. For small $\kappa$ the curves collapse onto a single $\kappa$-independent curve [Eq.~(\ref{eq:Q0}) below], while increasing $\kappa$ generally suppresses $Q_{\rm discharge}$. In the continuum regime $\kappa\ll1$, treating the charge number as a continuous variable gives, at fixed $\varepsilon$,
\begin{equation}
Q_{\rm discharge}
\mathop{\longrightarrow}_{\kappa\to0}
\frac{1}{2}k_{\rm B}T z^2(\varepsilon),
\label{eq:Q0}
\end{equation}
where $z(\varepsilon)=\sqrt{2}\,\mathrm{erfc}^{-1}(2\varepsilon)$; see Appendix~\ref{app:cont}. This expression, shown as the dashed curve in Fig.~\ref{fig:heat-eps}(a), agrees well with the exact results at small $\kappa$.

In Fig.~\ref{fig:heat-eps}(b), $Q_{\rm charge}$ approaches the Landauer cost as $\kappa$ becomes large, with oscillations at intermediate $\kappa$. By the argument given above for $\kappa\gg1$, Eq.~(\ref{eq:Qc}) gives
\begin{equation}
Q_{\rm charge}
\mathop{\longrightarrow}_{\kappa\to\infty}
k_{\rm B}T[\ln2-H(\varepsilon)].
\label{eq:Qc-limit}
\end{equation}
This Landauer cost gives the dashed curve in Fig.~\ref{fig:heat-eps}(b).

The oscillation of $Q_{\rm charge}$ seen in both Figs.~\ref{fig:heat}(b) and \ref{fig:heat-eps}(b) arises because the internal entropy of the discrete Gaussian, Eq.~(\ref{eq:peq}), depends on where its center $n_{\mathrm{i}}+\Delta n_{\rm BL}$, varied by $\kappa$ and $\varepsilon$, lies relative to the integer charge values. The internal entropy of the final state, Eq.~(\ref{eq:pfin}), is largest when $n_{\mathrm{i}}+\Delta n_{\rm BL}$ is near a half-integer, where the weight is shared almost equally between two adjacent charge states, and smallest when it is near an integer, where it concentrates on a single state. As $n_{\mathrm{i}}+\Delta n_{\rm BL}$ moves between integer and half-integer positions, $s(p^{\rm final})$---and hence $Q_{\rm charge}$---oscillates.

\begin{figure}[tb]
\includegraphics[width=\columnwidth]{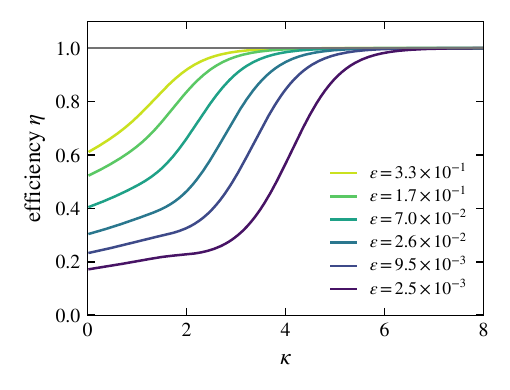}
\caption{\label{fig:eta}%
Energy efficiency $\eta$, Eq.~(\ref{eq:eta}), versus $\kappa$ at fixed error probabilities $\varepsilon$.
The efficiency increases monotonically with $\kappa$ and approaches the Landauer limit $\eta = 1$ (gray line) for $\kappa\gg1$.}
\end{figure}

\begin{figure}[tb]
\includegraphics[width=\columnwidth]{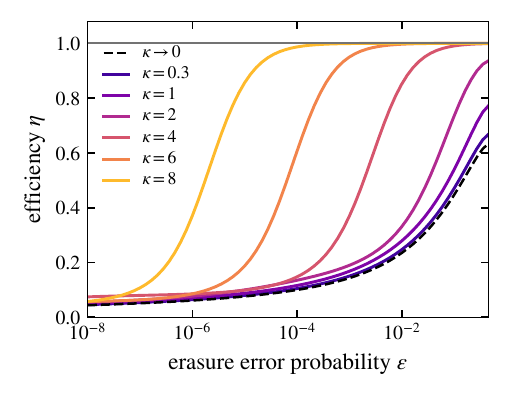}
\caption{\label{fig:eta-eps}%
Energy efficiency $\eta$ versus error probability $\varepsilon$ for several $\kappa$.
The black dashed curve shows the continuum-limit expression of Eq.~(\ref{eq:eta-cont}).
At fixed $\kappa$, $\eta$ increases with $\varepsilon$; as $\kappa$ grows the curves rise toward the Landauer limit $\eta=1$ (gray line).}
\end{figure}

\textit{Efficiency}---Combining the two heat contributions through Eq.~(\ref{eq:eta}) yields the efficiency shown in Fig.~\ref{fig:eta}.
For every $\varepsilon$, $\eta$ increases monotonically with $\kappa$, rising from a low value at $\kappa\ll1$ to unity at $\kappa\gg1$; at fixed $\kappa$ it is larger for larger $\varepsilon$.
The efficiency as a function of the error probability is shown in Fig.~\ref{fig:eta-eps}, and its two limiting behaviors follow from the heat expressions derived above.
For $\kappa\ll1$, $Q_{\mathrm{charge}}$ is negligible and $Q_{\mathrm{discharge}}\to\tfrac{1}{2}k_{\mathrm{B}}T z^{2}(\varepsilon)$ [Eq.~(\ref{eq:Q0})], so
\begin{equation}
\eta \xrightarrow{\;\kappa\to 0\;}
 \frac{2\left[\ln 2 - H(\varepsilon)\right]}{z^{2}(\varepsilon)},
\label{eq:eta-cont}
\end{equation}
which decreases for smaller $\varepsilon$, showing that a smaller error probability comes at the cost of efficiency~\cite{Shimizu2026,Freitas2022}; this continuum limit is the black dashed curve in Fig.~\ref{fig:eta-eps}.
For $\kappa\gg1$, $Q_{\mathrm{discharge}}\to 0$ while $Q_{\mathrm{charge}}\to k_{\mathrm{B}}T[\ln 2-H(\varepsilon)]$ [Eq.~(\ref{eq:Qc-limit})], so $Q/k_{\mathrm{B}}T\to\ln 2-H(\varepsilon)=-\Delta S$; hence the numerator and denominator of Eq.~(\ref{eq:eta}) coincide and
\begin{equation}
\eta \xrightarrow{\;\kappa\to\infty\;} 1
\label{eq:eta-tl}
\end{equation}
for every $\varepsilon$: the Landauer limit is attained.
Accordingly, at fixed $\kappa$ the efficiency in Fig.~\ref{fig:eta-eps} rises with $\varepsilon$, since a larger $\varepsilon$ reduces the required $\Delta n_{\mathrm{BL}}$ and hence $Q_{\rm discharge}$, and curves for larger $\kappa$ lie closer to $\eta=1$.

\textit{Discussion}---The physical origin of the recovery of the Landauer limit is the change of character of the initial state.
For any finite $\kappa$, the bimodal distribution of Eq.~(\ref{eq:pinit}) is not the equilibrium distribution of the single-well state function of Eq.~(\ref{eq:Psi}), and the mismatch is relaxed irreversibly in the discharge step.
For $\kappa \to \infty$, the initial state then coincides with the equilibrium state of the cell [Fig.~\ref{fig:scheme}(f)], the protocol becomes quasistatic from start to finish, and the DRAM cell becomes thermodynamically equivalent to the two-level systems in which the Landauer limit has been saturated.
The parameter $\kappa$ thus interpolates continuously between a multilevel system with a single-well state function, which is structurally barred from the Landauer limit, and an effective two-level memory, which attains it.

From a device perspective, $\kappa = e^{2}/(2Ck_{\mathrm{B}}T)$ is increased by reducing the storage capacitance or the operating temperature.
For the silicon single-electron device of Ref.~\cite{Shimizu2026} ($E_{\mathrm{c}} = 8.2$~meV), the effectively two-level regime $\kappa \gtrsim 4$---where the efficiency approaches the Landauer limit at a representative error probability $\varepsilon\approx2.6\times10^{-2}$---would be reached below roughly 25~K; alternatively, room-temperature operation at $\kappa \approx 4$ requires $E_{\mathrm{c}} \approx 100$~meV, i.e., a storage node of sub-attofarad capacitance, within reach of present single-electron technology~\cite{Nishiguchi2006}.
Our results therefore suggest that the aggressive downscaling of storage capacitors, usually motivated by density, is also the route along which the fundamental efficiency ceiling of charge-based memory rises toward the Landauer limit.

This conclusion does not imply that increasing $\kappa$ always improves the full operation of a DRAM device.
Beyond the erasure step analyzed here, a practical memory must also be read out.
In a conventional DRAM, readout is destructive and is performed by the discharge itself: turning on the wordline transistor shares the stored charge onto the much larger bitline capacitance, and the resulting small bitline-voltage change is amplified by a sense amplifier~\cite{Jacob2008,Ha2018} (distinct from the single-electron charge sensing used to probe the cell in Ref.~\cite{Shimizu2026}).
Since a larger $\kappa$ at fixed $\varepsilon$ narrows the charge distributions and reduces the separation between the two logical states, the charge available for this readout also shrinks (down to of order one electron), so increasing $\kappa$ trades a lower erasure heat against a smaller readout signal---a trade-off whose analysis, including the sense-amplifier process, we leave for future work.

\textit{Conclusion and outlook}---We have determined how $\kappa$ governs the energy efficiency of information erasure in a DRAM cell.
The nonquasistatic heat of the discharge step, which is responsible for the failure to reach the Landauer limit, is suppressed and eventually vanishes as $\kappa$ grows, while the quasistatic charge step approaches the Landauer cost; the efficiency increases monotonically and attains the Landauer limit once the cell becomes an effective two-level memory.
More fundamentally, these results identify $\kappa$ as the parameter that lifts the thermodynamic constraint imposed by the nonequilibrium initial state of volatile memory.

Several directions follow from this work.
The energy efficiency can be measured in the device of Ref.~\cite{Shimizu2026} at lower temperature, or in devices with smaller capacitance, to trace the curves of Fig.~\ref{fig:eta} as $\kappa$ is increased.
Our analysis is restricted to quasistatic driving, and extending it to finite-time protocols~\cite{Chen2026}, where additional dissipation arises from the finite operation speed, would connect the discreteness-controlled bound studied here to the finite-time costs of practical erasure.
Extensions to multibit architectures~\cite{Nishiguchi2006} and to other transistor--capacitor circuits such as static random-access memory (SRAM) cells and logic gates~\cite{Freitas2021,Yoshino2023} would further clarify how the competition between charging and thermal energies shapes the thermodynamic efficiency limits of electronic information-processing devices.

\begin{acknowledgments}
This work was supported by JSPS KAKENHI Grant Number JP26H02150.
OpenAI GPT-5.6 sol was used to generate an initial manuscript draft from author-prepared presentation slides, assist with subsequent author-directed revisions, and generate all plotting code used for the figures.
No AI-based image-generation tools were used.
All AI-assisted text, code, and resulting figures were reviewed and verified by the authors, who take full responsibility for the content of this work.
\end{acknowledgments}

\bibliography{references}

@article{Landauer1961,
  author  = {R. Landauer},
  title   = {Irreversibility and heat generation in the computing process},
  journal = {IBM J. Res. Dev.},
  volume  = {5},
  pages   = {183},
  year    = {1961},
  doi     = {10.1147/rd.53.0183},
}

@article{Bennett1982,
  author  = {C. H. Bennett},
  title   = {The thermodynamics of computation---a review},
  journal = {Int. J. Theor. Phys.},
  volume  = {21},
  pages   = {905},
  year    = {1982},
  doi     = {10.1007/BF02084158},
}

@article{Parrondo2015,
  author  = {J. M. R. Parrondo and J. M. Horowitz and T. Sagawa},
  title   = {Thermodynamics of information},
  journal = {Nat. Phys.},
  volume  = {11},
  pages   = {131},
  year    = {2015},
  doi     = {10.1038/nphys3230},
}

@article{Wolpert2019,
  author  = {D. H. Wolpert},
  title   = {The stochastic thermodynamics of computation},
  journal = {J. Phys. A},
  volume  = {52},
  pages   = {193001},
  year    = {2019},
  doi     = {10.1088/1751-8121/ab0850},
}

@article{Berut2012,
  author  = {A. B{\'e}rut and A. Arakelyan and A. Petrosyan and S. Ciliberto and R. Dillenschneider and E. Lutz},
  title   = {Experimental verification of {Landauer's} principle linking information and thermodynamics},
  journal = {Nature (London)},
  volume  = {483},
  pages   = {187},
  year    = {2012},
  doi     = {10.1038/nature10872},
}

@article{Jun2014,
  author  = {Y. Jun and M. Gavrilov and J. Bechhoefer},
  title   = {High-precision test of {Landauer's} principle in a feedback trap},
  journal = {Phys. Rev. Lett.},
  volume  = {113},
  pages   = {190601},
  year    = {2014},
  doi     = {10.1103/PhysRevLett.113.190601},
}

@article{Hong2016,
  author  = {J. Hong and B. Lambson and S. Dhuey and J. Bokor},
  title   = {Experimental test of {Landauer's} principle in single-bit operations on nanomagnetic memory bits},
  journal = {Sci. Adv.},
  volume  = {2},
  pages   = {e1501492},
  year    = {2016},
  doi     = {10.1126/sciadv.1501492},
}

@article{Yan2018,
  author  = {L. L. Yan and T. P. Xiong and K. Rehan and F. Zhou and D. F. Liang and L. Chen and J. Q. Zhang and W. L. Yang and Z. H. Ma and M. Feng},
  title   = {Single-atom demonstration of the quantum {Landauer} principle},
  journal = {Phys. Rev. Lett.},
  volume  = {120},
  pages   = {210601},
  year    = {2018},
  doi     = {10.1103/PhysRevLett.120.210601},
}

@article{Shimizu2026,
  author  = {T. Shimizu and K. Chida and G. Yamahata and K. Nishiguchi},
  title   = {Thermodynamic constraints in dynamic random-access memory cells: Experimental verification of energy efficiency limits in information erasure},
  journal = {Phys. Rev. Lett.},
  volume  = {136},
  pages   = {117103},
  year    = {2026},
  doi     = {10.1103/1sgm-dhys},
}

@article{Proesmans2020,
  author  = {K. Proesmans and J. Ehrich and J. Bechhoefer},
  title   = {Finite-time {Landauer} principle},
  journal = {Phys. Rev. Lett.},
  volume  = {125},
  pages   = {100602},
  year    = {2020},
  doi     = {10.1103/PhysRevLett.125.100602},
}

@article{Lee2022,
  author  = {J. S. Lee and S. Lee and H. Kwon and H. Park},
  title   = {Speed limit for a highly irreversible process and tight finite-time {Landauer's} bound},
  journal = {Phys. Rev. Lett.},
  volume  = {129},
  pages   = {120603},
  year    = {2022},
  doi     = {10.1103/PhysRevLett.129.120603},
}

@article{VanVu2023,
  author  = {T. {Van Vu} and K. Saito},
  title   = {Thermodynamic unification of optimal transport: Thermodynamic uncertainty relation, minimum dissipation, and thermodynamic speed limits},
  journal = {Phys. Rev. X},
  volume  = {13},
  pages   = {011013},
  year    = {2023},
  doi     = {10.1103/PhysRevX.13.011013},
}

@article{Zhen2021,
  author  = {Y.-Z. Zhen and D. Egloff and K. Modi and O. Dahlsten},
  title   = {Universal bound on energy cost of bit reset in finite time},
  journal = {Phys. Rev. Lett.},
  volume  = {127},
  pages   = {190602},
  year    = {2021},
  doi     = {10.1103/PhysRevLett.127.190602},
}

@article{Scandi2022,
  author  = {M. Scandi and D. Barker and S. Lehmann and K. A. Dick and V. F. Maisi and M. Perarnau-Llobet},
  title   = {Minimally dissipative information erasure in a quantum dot via thermodynamic length},
  journal = {Phys. Rev. Lett.},
  volume  = {129},
  pages   = {270601},
  year    = {2022},
  doi     = {10.1103/PhysRevLett.129.270601},
}

@article{Oikawa2025,
  author  = {S. Oikawa and Y. Nakayama and S. Ito and T. Sagawa and S. Toyabe},
  title   = {Experimentally achieving minimal dissipation via thermodynamically optimal transport},
  journal = {Nat. Commun.},
  volume  = {16},
  pages   = {10424},
  year    = {2025},
  doi     = {10.1038/s41467-025-66519-9},
}

@article{Freitas2021,
  author  = {N. Freitas and J.-C. Delvenne and M. Esposito},
  title   = {Stochastic thermodynamics of nonlinear electronic circuits: A realistic framework for computing around $kT$},
  journal = {Phys. Rev. X},
  volume  = {11},
  pages   = {031064},
  year    = {2021},
  doi     = {10.1103/PhysRevX.11.031064},
}

@article{Freitas2022,
  author  = {N. Freitas and K. Proesmans and M. Esposito},
  title   = {Reliability and entropy production in nonequilibrium electronic memories},
  journal = {Phys. Rev. E},
  volume  = {105},
  pages   = {034107},
  year    = {2022},
  doi     = {10.1103/PhysRevE.105.034107},
}

@article{Yoshino2023,
  author  = {D. Yoshino and Y. Tokura},
  title   = {Thermodynamics of computation for {CMOS} {NAND} gate},
  journal = {J. Phys. Soc. Jpn.},
  volume  = {92},
  pages   = {124004},
  year    = {2023},
  doi     = {10.7566/JPSJ.92.124004},
}

@article{Nishiguchi2014,
  author  = {K. Nishiguchi and Y. Ono and A. Fujiwara},
  title   = {Single-electron thermal noise},
  journal = {Nanotechnology},
  volume  = {25},
  pages   = {275201},
  year    = {2014},
  doi     = {10.1088/0957-4484/25/27/275201},
}

@article{Nishiguchi2006,
  author  = {K. Nishiguchi and A. Fujiwara and Y. Ono and H. Inokawa and Y. Takahashi},
  title   = {Room-temperature-operating data processing circuit based on single-electron transfer and detection with metal-oxide-semiconductor field-effect transistor technology},
  journal = {Appl. Phys. Lett.},
  volume  = {88},
  pages   = {183101},
  year    = {2006},
  doi     = {10.1063/1.2200475},
}

@book{Jacob2008,
  author  = {B. Jacob and S. W. Ng and D. T. Wang},
  title   = {Memory Systems: Cache, {DRAM}, Disk},
  publisher = {Morgan Kaufmann/Elsevier},
  address = {Burlington, MA},
  year    = {2008},
}

@phdthesis{Ha2018,
  author  = {H. Ha},
  title   = {Understanding and improving the energy efficiency of {DRAM}},
  school  = {Stanford University},
  year    = {2018},
  note    = {\url{https://purl.stanford.edu/yp843xn4828}},
}

@article{Chen2026,
  author  = {Songela W. Chen and David T. Limmer},
  title   = {Optimal Control of Bit Erasure in Stochastic Random Access Memory},
  journal = {PRX Energy},
  volume  = {5},
  pages   = {023011},
  year    = {2026},
  doi     = {10.1103/bmsv-mlq5},
}

\appendix

\section{Initial distribution across the crossover}
\label{app:pinit}

\begin{figure}[t]
\centering
\includegraphics[width=\linewidth]{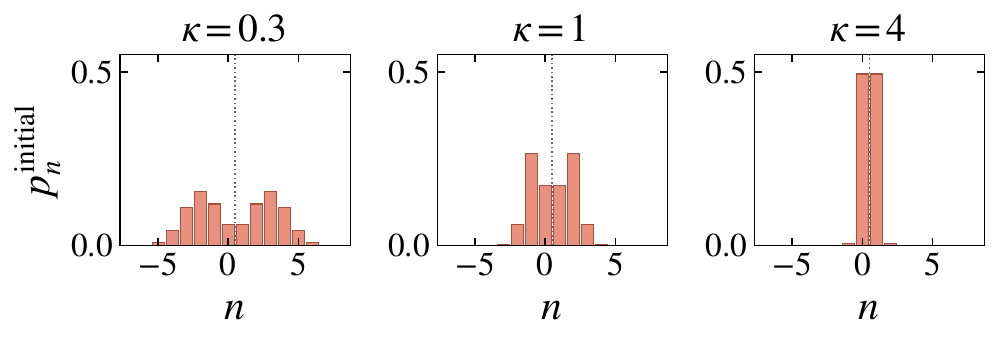}
\caption{\label{fig:pinit-kappa}%
Initial distribution $p_n^{\mathrm{initial}}$, Eq.~(\ref{eq:pinit}) of the main text, at fixed erasure error probability $\varepsilon=2.6\times10^{-2}$ for $\kappa=0.3$, $1$, and $4$. Vertical dotted lines mark $n=n_{\mathrm{i}}=0.5$. As $\kappa$ increases, the distribution narrows and concentrates on $n=0$ and $n=1$.}
\end{figure}

Figure~\ref{fig:pinit-kappa} shows the initial distribution $p_n^{\mathrm{initial}}$, Eq.~(\ref{eq:pinit}) of the main text, at fixed error probability $\varepsilon$ for three values of $\kappa$.
For small $\kappa$ the distribution is broad and spreads over many charge states, with the two components centered at $n_{\mathrm{i}}\pm\Delta n_{\mathrm{BL}}$ well separated from the discharge center $n_{\mathrm{i}}$.
As $\kappa$ increases, each component narrows and the required displacement $\Delta n_{\mathrm{BL}}$ shrinks, so the two components merge toward $n=0$ and $n=1$.
For $\kappa\gg1$ the weight is concentrated almost entirely on these two states; the small residual weight on $n\le 0$ is the erasure error probability $\varepsilon$, which is held fixed across the panels. In this limit the initial distribution nearly coincides with the discharged distribution $p^{\rm discharged}=p_n^{\rm eq}(n_{\mathrm{i}})$, so the discharge step induces almost no relaxation and $Q_{\rm discharge}\to0$.

\section{Continuum-limit derivation of the discharge heat}
\label{app:cont}

Here we derive the $\kappa\to 0$ discharge heat, Eq.~(\ref{eq:Q0}) of the main text.
In the continuum regime $\kappa\ll 1$, many charge states are populated and the discrete distribution $p_n^{\mathrm{eq}}(n_{\mathrm{BL}})$ is well approximated by the continuous Gaussian
\begin{equation}
g_x(n_{\mathrm{BL}})
=\frac{\exp[-(x-n_{\mathrm{BL}})^2/(2\sigma^2)]}{\sqrt{2\pi}\,\sigma},
\label{eq:gdef}
\end{equation}
so that sums over $n$ may be replaced by integrals over $x$.
Here $\sigma=1/\sqrt{2\kappa}$ is the width introduced in the main text, so that $\sigma^2=k_{\mathrm{B}}T/(2E_{\mathrm{c}})$.
The discharge step is carried out at $n_{\mathrm{BL}}=n_{\mathrm{i}}=0.5$, where $\Psi_n(n_{\mathrm{i}})=E_{\mathrm{c}}(n-n_{\mathrm{i}})^2$.
Define the expectation value of the state function for a distribution $p_n$ as
\begin{equation}
\langle\Psi\rangle_p \equiv \sum_n p_n\Psi_n(n_{\mathrm{i}}).
\label{eq:psiavg}
\end{equation}
Then, for an equilibrium distribution centered at $n_{\mathrm{BL}}$, in the continuum limit,
\begin{align}
\langle\Psi\rangle_{p^{\mathrm{eq}}(n_{\mathrm{BL}})}
&\to E_{\mathrm{c}}\int_{-\infty}^{\infty} dx\,g_x(n_{\mathrm{BL}})(x-n_{\mathrm{i}})^2 \nonumber\\
&=E_{\mathrm{c}}\left[\sigma^2+(n_{\mathrm{BL}}-n_{\mathrm{i}})^2\right].
\label{eq:moment}
\end{align}

Using Eq.~(\ref{eq:pinit}) of the main text, the initial state is the average of the distributions centered at $n_{\mathrm{i}}-\Delta n_{\mathrm{BL}}$ and $n_{\mathrm{i}}+\Delta n_{\mathrm{BL}}$, so Eq.~(\ref{eq:moment}) gives
\begin{align}
\langle\Psi\rangle_{p^{\mathrm{initial}}}
&=\frac{1}{2}\langle\Psi\rangle_{p^{\mathrm{eq}}(n_{\mathrm{i}}-\Delta n_{\mathrm{BL}})}
 +\frac{1}{2}\langle\Psi\rangle_{p^{\mathrm{eq}}(n_{\mathrm{i}}+\Delta n_{\mathrm{BL}})} \nonumber\\
&\to E_{\mathrm{c}}\left(\sigma^2+\Delta n_{\mathrm{BL}}^2\right).
\label{eq:Psi-init}
\end{align}
Similarly, since $p^{\mathrm{discharged}}=p^{\mathrm{eq}}(n_{\mathrm{i}})$,
\begin{equation}
\langle\Psi\rangle_{p^{\mathrm{discharged}}}
=\langle\Psi\rangle_{p^{\mathrm{eq}}(n_{\mathrm{i}})}
\to E_{\mathrm{c}}\sigma^2.
\label{eq:Psi-dis}
\end{equation}
Substituting Eqs.~(\ref{eq:Psi-init}) and (\ref{eq:Psi-dis}) into Eq.~(\ref{eq:Qd}) of the main text yields
\begin{equation}
Q_{\mathrm{discharge}}
=\langle\Psi\rangle_{p^{\mathrm{initial}}}
 -\langle\Psi\rangle_{p^{\mathrm{discharged}}}
\to E_{\mathrm{c}}\Delta n_{\mathrm{BL}}^2.
\label{eq:Qd-Dn}
\end{equation}

It remains to express $\Delta n_{\mathrm{BL}}$ in terms of the erasure error probability.
In the continuum approximation, the boundary between the logical states $n\le0$ and $n\ge1$ is placed at their midpoint $n_{\mathrm{i}}=0.5$.
For the final erase-to-1 distribution centered at $n_{\mathrm{i}}+\Delta n_{\mathrm{BL}}$, the weight on the ``0'' side is obtained by introducing $y=[x-(n_{\mathrm{i}}+\Delta n_{\mathrm{BL}})]/(\sqrt{2}\sigma)$:
\begin{align}
\varepsilon
&=\int_{-\infty}^{n_{\mathrm{i}}} dx\,g_x(n_{\mathrm{i}}+\Delta n_{\mathrm{BL}}) \nonumber\\
&=\frac{1}{\sqrt{\pi}}
\int_{-\infty}^{-\Delta n_{\mathrm{BL}}/(\sqrt{2}\sigma)}
 dy\,e^{-y^2} \nonumber\\
&=\frac{1}{\sqrt{\pi}}
\int_{\Delta n_{\mathrm{BL}}/(\sqrt{2}\sigma)}^\infty
 dy\,e^{-y^2} \nonumber\\
&=\frac{1}{2}\,\mathrm{erfc}\!\left(
  \frac{\Delta n_{\mathrm{BL}}}{\sqrt{2}\sigma}
 \right).
\label{eq:eps-tail}
\end{align}
Here we used the evenness of $e^{-y^2}$ and the definition
$\mathrm{erfc}(z)=(2/\sqrt{\pi})\int_z^\infty e^{-t^2}\,dt$.
Thus
\begin{equation}
\Delta n_{\mathrm{BL}}=z(\varepsilon)\sigma,
\qquad
z(\varepsilon)=\sqrt{2}\,\mathrm{erfc}^{-1}(2\varepsilon),
\label{eq:D-scaling-app}
\end{equation}
which gives the fixed-$\varepsilon$ scaling $\Delta n_{\mathrm{BL}}\propto\sigma\propto\kappa^{-1/2}$.
Combining Eqs.~(\ref{eq:Qd-Dn}) and (\ref{eq:D-scaling-app}) with $\sigma^2=k_{\mathrm{B}}T/(2E_{\mathrm{c}})$ gives
\begin{equation}
Q_{\mathrm{discharge}}
\to E_{\mathrm{c}}z^2(\varepsilon)\sigma^2
=\frac{1}{2}k_{\mathrm{B}}Tz^2(\varepsilon),
\end{equation}
which is Eq.~(\ref{eq:Q0}) of the main text.

\end{document}